# Digitize Your Body and Action in 3-D at Over 10 FPS: Real Time Dense Voxel Reconstruction and Marker-less Motion Tracking via GPU Acceleration

Jian Song*, Yatao Bian*[1], Junchi Yan, Xu Zhao, Yuncai Liu

Department of Automation, Shanghai Jiao Tong University, and Key Laboratory of System Control and Information Processing, Ministry of Education of China, Shanghai 200240, China

**Abstract**

In this paper, we present an approach to reconstruct 3-D human motion from multi-cameras and track human skeleton using the reconstructed human 3-D point (voxel) cloud. We use an improved and more robust algorithm, *probabilistic shape from silhouette* to reconstruct human voxel. In addition, the annealed particle filter is applied for tracking, where the measurement is computed using the reprojection of reconstructed voxel. We use two different ways to accelerate the approach. For the CPU only acceleration, we leverage Intel TBB to speed up the hot spot of the computational overhead and reached an accelerating ratio of 3.5 on a 4-core CPU. Moreover, we implement an intensively paralleled version via GPU acceleration without TBB. Taking account all data transfer and computing time, the GPU version is about *400* times faster than the original CPU implementation, leading the approach to run at a real-time speed.

## 1 Introduction

Visual information based human motion analysis is an important research direction in computer vision. And 3-D human motion reconstruction and tracking is one of the most challenging tasks. There are huge applications of the related techniques, ranging from virtual realization, film animation, intelligent surveillance, high level human machine interaction, video transmission and compression to sports and medical treatment. With the rapid development of hardware facilities and the theory of computer vision, currently, building a video based human motion construction system for real application is attracting more and more attentions from both academic and industrial fields. Considering the intensive computational demand, GPU will definitely play a very important role in this campaign. This is also the reason why we make our best to take part in this contest.

The main task of 3-D human motion reconstruction is to estimate the motion parameters and body configuration of 3-D human body model, by utilizing visual information from multiple cameras. It also means that we aim to get the position and the configurations of human body parts in real 3-D world. Due to the complexity of this problem, the computational complexities of related algorithms are impermissible to a real-time implementation in CPU. Therefore, the algorithm acceleration using GPU is a nice choice. Nowadays, GPU is getting its wide recognition and application among scholars, especially in the 3-D reconstruction field. A latest relevant case is the work presented by Jan-Michael Frahm et al. from UNC, who published their work on ECCV (European Conference on Computer Vision) [3]. They collected the photos from internet (millions of images), thanks to GPU for acceleration, they successfully reconstruct the city of Rome in one day, without the help of cloud computing. In comparison, Sameer Agarwal from University of Washington achieve such performance in their ICCV (International Conference on Computer Vision) paper, owing to the high performance from cloud computing (62 workstations), and the number of images they approach is only 150,000, significantly fewer than the work [3]. From above cases, one can get a clear perspective that GPU is playing very important role in computer vision field.

---

[1] * denotes equal contribution

In our project, there are two classes of core techniques: **3-D voxel reconstruction of human body** and **3-D body pose tracking**. Both of them are time consuming. For 3-D voxel reconstruction, the main algorithm is Shape From Silhouette (SFS). Regarding with SFS techniques, Sofiane Yous et al. leveraged the GPU's advantage in light tracking, and achieved the real time 3-D reconstruction in some published datasets [5] in 2007. Moreover, Neal Orman et al implemented a free viewpoint rendering system thanks to the acceleration capability brought by GPU [6]. Compared with the traditional SFS methods, here we adopt the method proposed in [1][7] to achieve better robustness, which is insensitive to outliers, while at the price of computational inefficiency. And since we have not found any GPU based implementation of this method, we are exploring the potential high efficiency when GPU is adopted in this project.

After obtaining the 3-D voxel of human body, the next step is to perform human pose estimation and tracking. This is another time-consuming part of this project. In industrial field, 4D View Solution[2] works on 3-D pose estimation but mainly using parallel CPU. In our project, we select GPU as the accelerator. The core algorithm for pose tracking is Annealed Particle Filter (APF). Regarding with the APF based tracking approach, inherently it is a particle filter algorithm, which has a natural form of parallel computing and therefore it is expected to have a high potential in acceleration.

Towards the aim of 3-D human body voxel reconstruction and 3-D pose tracking, the presented project bears the following aspects:

- Using multiple cameras to obtain visual information of human motion in indoor environments.
- Using human skeleton model as the prior model of human body.
- Achieving generality to different types of human motion.
- Achieving quick and robust solution to 3-D human pose tracking.

In sum, the main algorithms in this project are listed as follows:

- For 3-D human body reconstruction, probabilistic shape from silhouette approach is implemented [1][7].
- For 3-D human pose tracking, annealed particle filter method is used [2].

---

[2] A start-up company is born out of Rhone-Alpes National Institute for Research in Computer Science and Control. http://r24085.ovh.net/company.html

We next will introduce these algorithms in detail.

## 2 Algorithm Design

### 2.1 System Flowchart

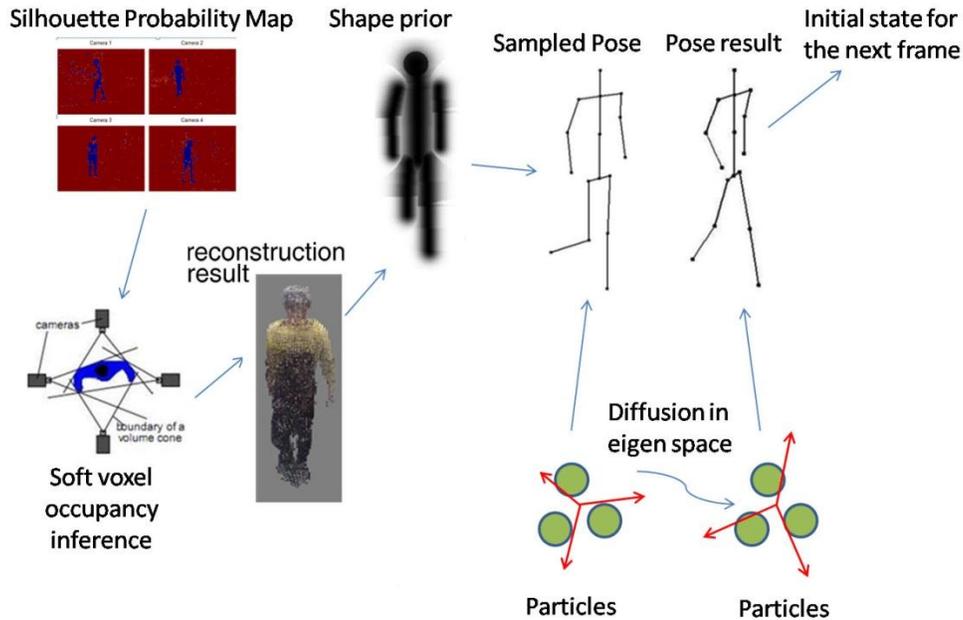

Figure 1 System flowchart

Figure 1 shows the system flowchart. For a given image, we first use *probabilistic shape from silhouette* (PSFS) [1][7] to get reconstructed voxel of human surface. Then with the information from reconstructed human body, we track human skeleton in an annealed particle filter [2] framework. We will give detailed descriptions of the two algorithm parts in the following sub chapters.

### 2.2 3-D Voxel Reconstruction

#### 2.2.1 General Reconstruction Methods

Usually the two main methods used for reconstruction are based on silhouette and photometry. Photometry-based methods [9][10][11] are sensitive to camera registrations and have a quite high complexity. Silhouette-based methods are popular for use in multi-camera environments mainly due to their simplicity and computational efficiency.

Shape from silhouette [12][13]is a silhouette-based method that is well studied. It first gets the silhouette using a simple background subtraction method. Figure 2 shows how shape from silhouette works. As perspective projection suggests, the object in the scene should be located inside the three dimensional centrum formed by the camera center and silhouette. It is obvious that an object can be reconstructed by carving a big region of interest with multi-view images taken from that object.

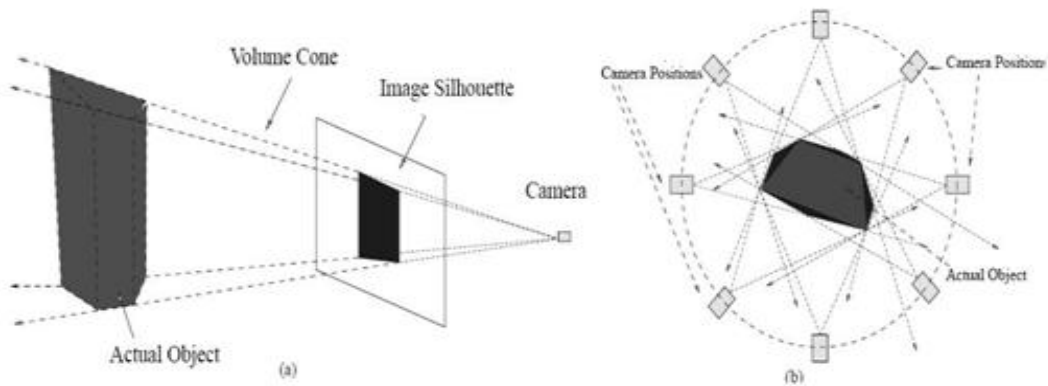

Figure 2 Generalization of 3-D Reconstruction algorithm [12]

### 2.2.2 Improved Scheme: Probabilistic Shape from Silhouette

We improved the original shape from silhouette. We use the visual hull of the human body as a basis of our algorithm. While very often silhouettes are used to infer shapes in a two-step process: an individual decision about silhouette occupancy is made on a per-view basis, and then shape is inferred geometrically from all available silhouettes using visual hull methods; yet here we infer the probability of each voxel occupancy by fusing the information from image cues in an incremental updating Bayesian framework as first proposed in the work [7] and further improved by our work [1]. The following algorithm is from our published paper [1].

In order to make full use of the image information from multi-views, we do not segment the binary silhouette of each image on the fly. Instead, we adopt a Bayesian framework to calculate the silhouette probability of each pixel. For a pixel $p$ in image $r$, the posterior probability of representing foreground can be calculated by Bayesian theory:

$$P(\mathcal{F}_r^p = 1 | I_r^p) = \frac{P(I_r^p | \mathcal{F}_r^p = 1) P(\mathcal{F}_r^p = 1)}{\sum_{\mathcal{F}_r^p = 0}^{1} P(I_r^p | \mathcal{F}_r^p) P(\mathcal{F}_r^p)} \quad (1)$$

where $\mathcal{F}_r^p = 1$ indicates the pixel $p$ in image $r$ represents foreground, otherwise, $\mathcal{F}_r^p = 0$. $I_r^p$ denotes the color feature vector of pixel $p$ in image $r$. For simplicity, we model $P(I_r^p | \mathcal{F}_r^p = 0) \sim N(I_r^p | \mu_r^p, \sigma_r^p)$, a single Gaussian model. Since the foreground color information is unknown for the first frame, we assume the foreground obeys a uniform distribution:

$P(I_r^p | \mathcal{F}_r^p = 1) \sim U(I_r^p)$ We choose not to favor any pixel location by setting the prior term $P(\mathcal{F}_r^p = 1) = P(\mathcal{F}_r^p = 0) = 0.5$. And we obtain the silhouette likelihood map defined as:

$$SLM_r^p = P(\mathcal{F}_r^p = 1 | I_r^p) \quad (2)$$

Figure 3 shows silhouette likelihood map of four views from tow sequences. There are many holes and noises which cannot be removed using normal shape from silhouette method which results not perfect reconstructed voxels.

Consequently, given a set of images $\{I^P\}_n = \{I_1^P, I_2^P, ..., I_n^P\}$, one can calculate the silhouette likelihood maps $\{S^P\}_n = \{S_1^P, S_2^P, ..., S_n^P\}$.

Similar to the process of obtaining silhouette, given the silhouette likelihood map of multiple images $\{S^P\}_n = \{S_1^P, S_2^P, ..., S_n^P\}$, we estimate the posterior probability representing occupancy of each voxel in volume of interest as follows:

$$P(V_i = 1 | \{S^P\}_n) = \frac{P(\{S^P\}_n | V_i = 1) P(V_i = 1)}{\sum_{V_i=0}^{1} P(\{S^P\}_n | V_i) P(V_i)} \quad (3)$$

where $V_i = 1$ denotes the voxel $i$ is occupied by an object, otherwise, $V_i = 0$. The pixel of the silhouette likelihood map $\{S^P\}_n = \{S_1^P, S_2^P, ..., S_n^P\}$ s the projection of voxel $V_i$. The term $P(\{S^P\}_n | V_i = 1)$ in Eq.(3) can be calculated in Eq. (4):

$$\begin{aligned} P(\{S^P\}_n | V_i) &= P(\{S^P\}_{n-1} | V_i) P(S_n^P | V_i) \\ &= \Pi_{k=1}^{n} P(S_k^P | V_i) \end{aligned} \quad (4)$$

Then the problem of voxel occupancy inference is reduced to model the term $P(S_k^P | V_i) (k = 1, 2, ..., n)$ Here we introduce a latent variable $O_k^P$. $O_k^P = 1$ denotes there exits object occluding the voxel $V_i$ on the viewing line connecting the voxel and its projection $S_k^P$, otherwise $O_k^P = 0$. Thus we get the following equation:

$$P(S_k^P | V_i) = \sum_{O_k^P} P(S_k^P | O_k^P, V_i) P(O_k^P) \quad (5)$$

For the term $P(O_k^P)$, we set $P(O_k^P = 0) = P(O_k^P = 1) = 0.5$ indicating with no prior preference; for the term $P(S_k^P | O_k^P = 0, V_i = 0)$ which indicates that there exist neither voxel $V_i$ nor other object lying on the viewing line of pixel $S_k^P$, we have:

$$P(S_k^P | O_k^P = 0, V_i = 0) = 1 - SLM_k^P \quad (6)$$

For $P(S_k^P | O_k^P = 1, V_i = 0), P(S_k^P | O_k^P = 0, V_i = 1), P(S_k^P | O_k^P = 1, V_i = 1)$, which denote there at least exists one voxel lying on the viewing line, we assume:

$$P(S_k^P | O_k^P = 1, V_i = 0) = SLM_k^P \quad (7)$$

$$P(S_k^p \mid O_k^p = 0, V_i = 1) = SLM_k^p \tag{8}$$

$$P(S_k^p \mid O_k^p = 1, V_i = 1) = SLM_k^p \tag{9}$$

Finally we filter the probability of voxels occupied by human body and then perform a thresholding process, to remove voxels inside human body and get the surface voxels.

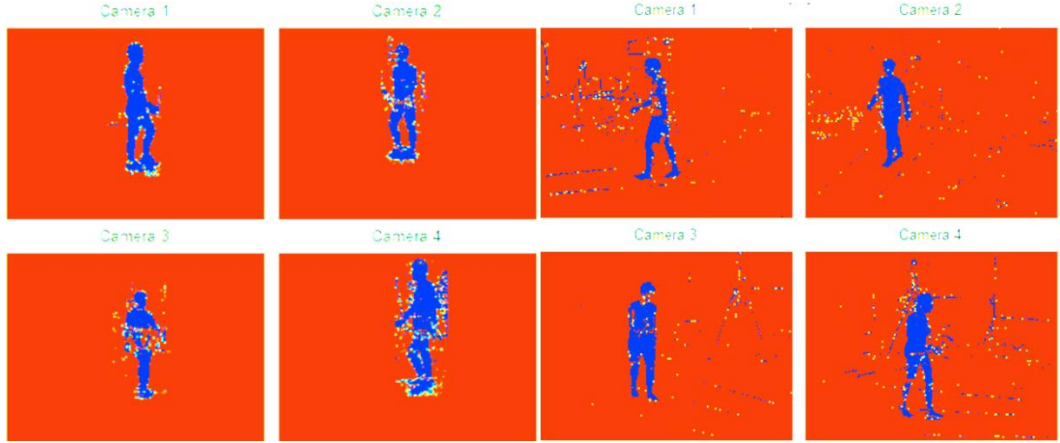

Figure 3 Silhouette likelihood map
a) from our dataset        b) from Brown University's dataset

## 2.3 Human Skeleton Tracking

### 2.3.1 Annealed particle filter

Particle Filter (PF) is based on Monte Carlo methods. It represents the probability distribution by particle set which can be used in any state space model. The main idea of Particle filter is to select random state particle from posteriori. This process is a sequential important sampling. In other words, PF is an Approximation Method for probability distribution function by finding a set of random sample in state space. Although the probability distribution is only an approximate for real distribution, it drops the limitation of random variables satisfying Gaussian distribution in nonlinear filter because of the non-parameter structure. Therefore the particle filter has wider and stronger modeling ability. The procedure of Particle Filter can be seen from Figure 4.

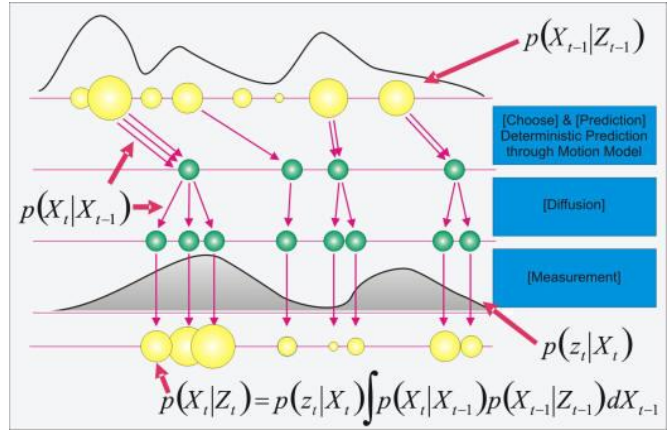
Figure 4 Particle filter iteration loop

**Simulated annealing**

Simulated annealing is an approach for minimizing multivariate functions. The term simulated annealing derives from the roughly analogous physical process of heating and then slowly cooling a substance to obtain a strong crystalline structure.

The cooling schedule is to anneal the problem from a random solution to a good, frozen, placement. Specifically, we need a starting hot temperature (or a heuristic for determining a starting temperature for the current problem) and rules to determine when the current temperature should be lowered, by how much the temperature should be lowered, and when annealing should be terminated.

**Annealed particle filter**

The process of Annealed particle filter (APF) [14] is stated as follow:

1. For every time step $T_k$, an annealing run is started at layer M, with m = M.

2. Each layer of an annealing run is initialized by a set of un-weighted particles $S_{k,m}$.

3. Each of these particles is then assigned a weight

$$\pi_{k,m}^i \propto \omega_m(Z_k, s_{k,m}^i) \tag{10}$$

which are normalized so that $\sum_N \pi_{k,m}^i = 1$, The set of weighted particles $S_{k,m}^\pi$ has now been formed.

4. N particles are drawn randomly from $S_{k,m}^\pi$ with replacement and with a probability equal to their weighting $\pi_{k,m}^i$. As the $n^{th}$ particle is chosen it is used to produce the particle $s_{k,m-1}^n$ using

$$s_{k,m-1}^n = s_{k,m}^n + B_m \tag{11}$$

where $B_m$ is a multi-variant Gaussian random variable with variance $P_m$ and mean 0.

5. The set $S_{k,m-1}$ has now been produced which can be used to initialize layer m - 1. The process is repeated until we arrive at the set $S_{k,0}^{\pi}$

6. $S_{k,0}^{\pi}$ is used to estimate the optimal model configuration $X_k$ using

$$X_k = \sum_{i=1}^{N} s_{k,0}^i \pi_{k,0}^i \tag{12}$$

7. The set $S_{k+1,M}$ is then produced from using $S_{k,0}^{\pi}$

$$s_{k+1,M}^n = s_{k,0}^n + B_0 \tag{13}$$

This set is then used to initialize layer M of the next annealing run at $t_{k+1}$.

### 2.3.2 Tracking Skeleton using Reconstructed Voxels

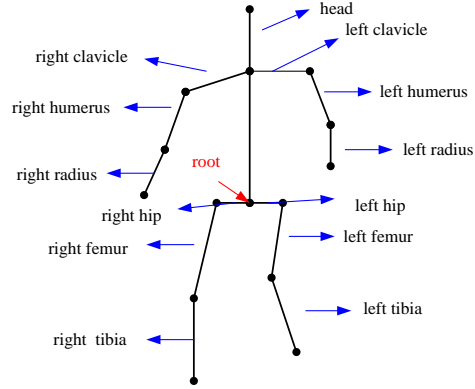

Figure 5 Human skeleton model

10 cylinders are set for different part of human body. The following table shows the corresponding part of the cylinders

| 1 | TORSO |
|---|---|
| 2 | Left thigh (upper leg) |
| 3 | Left calf (lower leg) |
| 4 | Right thigh (upper leg) |
| 5 | Right calf (lower leg) |
| 6 | Left upper arm |
| 7 | Left lower arm |
| 8 | Right upper arm |
| 9 | Right lower arm |
| 10 | Head |

Table 1 10 cylinders of human

We can get articular human model by setting base and bottom radius and length for each cylinder. This model has 31 degrees of Freedom, representing position and freedom of the ten cylinders.

Considering the 31 DOF of skeleton in one frame as a particle vector, and using the projection of voxels to be the prior knowledge, we can find an optimized particle and the get the new skeleton in the annealed particle filter framework. First we propagate the particles in the 31 dimension space. Each of the particles represents one pose of a skeleton. Then we match images generated by these particles and the projection of voxels in 2-D images and also their edges. We use the matched value as weight numbers for particles and we get a new particle after adding all these weighted particles. This new particle is then pushed into the next annealing layer. After propagating particles in all layers, we get the optimized particle, which represents the human skeleton in a new frame. Particle number is usually set to a value from tens to hundreds depending on the tracking condition. And the annealing layer is usually set to around 10.

**Particle Weight**
Weight value in formula (10) is divided into two parts. One is from the edge information,

$$\sum\nolimits^e (X, Z) = \frac{1}{N} \sum_{i=1}^{N} (1 - p_i^e(X, Z)) \tag{14}$$

And the other is from the silhouette,

$$\sum\nolimits^r (X, Z) = \frac{1}{N} \sum_{i=1}^{N} (1 - p_i^r(X, Z)) \tag{15}$$

While we have multi-cameras in reconstruction and tracking, the final weight is a sum of that two parts from all cameras, that is

$$\omega(X, Z) = \exp - (\sum_{i=1}^{C} (\sum\nolimits_i^e (X, Z) + \sum\nolimits_i^r (X, Z))) \tag{16}$$

Where C indicates the camera number

## 3  Algorithm Parallelization

### 3.1  Parallelism Analysis
Our scheme is based on computing and processing of images (digitized as matrix in practice), which has high parallelism to the task itself.

From the micro point of view, many intermediate processes are math operations of images or matrices. In the algorithm, large scale data is stored as continuous large array (2-dimensions or 3-demensions). These math operations can be accelerated by GPU parallelization.

From the macro point of view, both reconstruction and tracking parts in our algorithm are the procedures of processing multi-camera information. These separate processes of each camera can be accelerated by parallelization.

The specific analysis is presented in the following sub chapters according to the nature character of reconstruction and tracking algorithm.

### 3.1.1   3-D Voxel Reconstruction

The hot spot of 3-D reconstruction is the computation of voxel. And the color rendering is also an iterative process of all voxels.

Some parallelizable parts of the algorithm are listed as follows:

- Reconstruction problem is the procedure of analyzing 3-D space information after image processing, the data of which is multi-dimension matrix whose math operations has parallelism.
- Both reconstruction and color rendering are superposition of processing procedures of multi-cameras which are independent to each other and lead to parallelism.
- Reconstruction is the procedure of filtering all the voxels in three dimension region, the procedures for each voxel has nothing to do with others while the number of voxel is very large. Camera image information is processed in this procedure meanwhile. Great acceleration can be achieved via parallelization of voxel and image pixels processing.
- Color rendering aims to get color information by processing the reconstructed voxels. The large number of voxel and independence of each procedure for one voxel will largely contribute to acceleration via parallelization.

### 3.1.2   Human Skeleton Tracking

The hot spot of human skeleton tracking lies in the weight assigning process.

Some parallelizable parts of the algorithm are listed as follows:

- Tracking problem is the procedure of analyzing matching degree between real world images and constructed images after image processing, the data of which is multi-dimension matrix whose math operations has parallelism.
- The particles' weights are computed by formula(10), which will have a significant acceleration due to the large number of particles and the independence of each weight computing procedure.
- In the procedure of assigning weight for one specific particle, formula(16) is a superposition process from several cameras. And the independence of each camera's processing leads to parallelism.

## 3.2   Parallellization on Multi-Core CPU

To parallelize our program via GPU, we first parallelized it using multi-threads on multi-core CPU.

The parallelism on CPU is different with that on GPU since CPUs cannot hold as many threads as GPU running at the same time and the thread control is done explicitly. So we only paralleled the top level of all parallelizable parts.

We use Intel® Threading Building Blocks（Licensed under GPLv2 with the runtime exception） to build parallel programs on CPU.

In the voxel reconstruction part, we parallelized the calculating process of each camera while constructing voxels and rendering colors. And in the human skeleton tracking part, we parallelized the time-consuming process of each particle. These paralleled program parts are represented under the name of XXX_Invoker.

## 3.3 Parallelization on GPU

### 3.3.1 Algorithm Restructuring

GPU and CPU are adept in different computational demand scenarios because of architecture differences. Taking into account these differences and in accordance with the GPU computing features, we refined the algorithm structure. The overall frameworks of CPU and GPU algorithms are shown in the following figure.

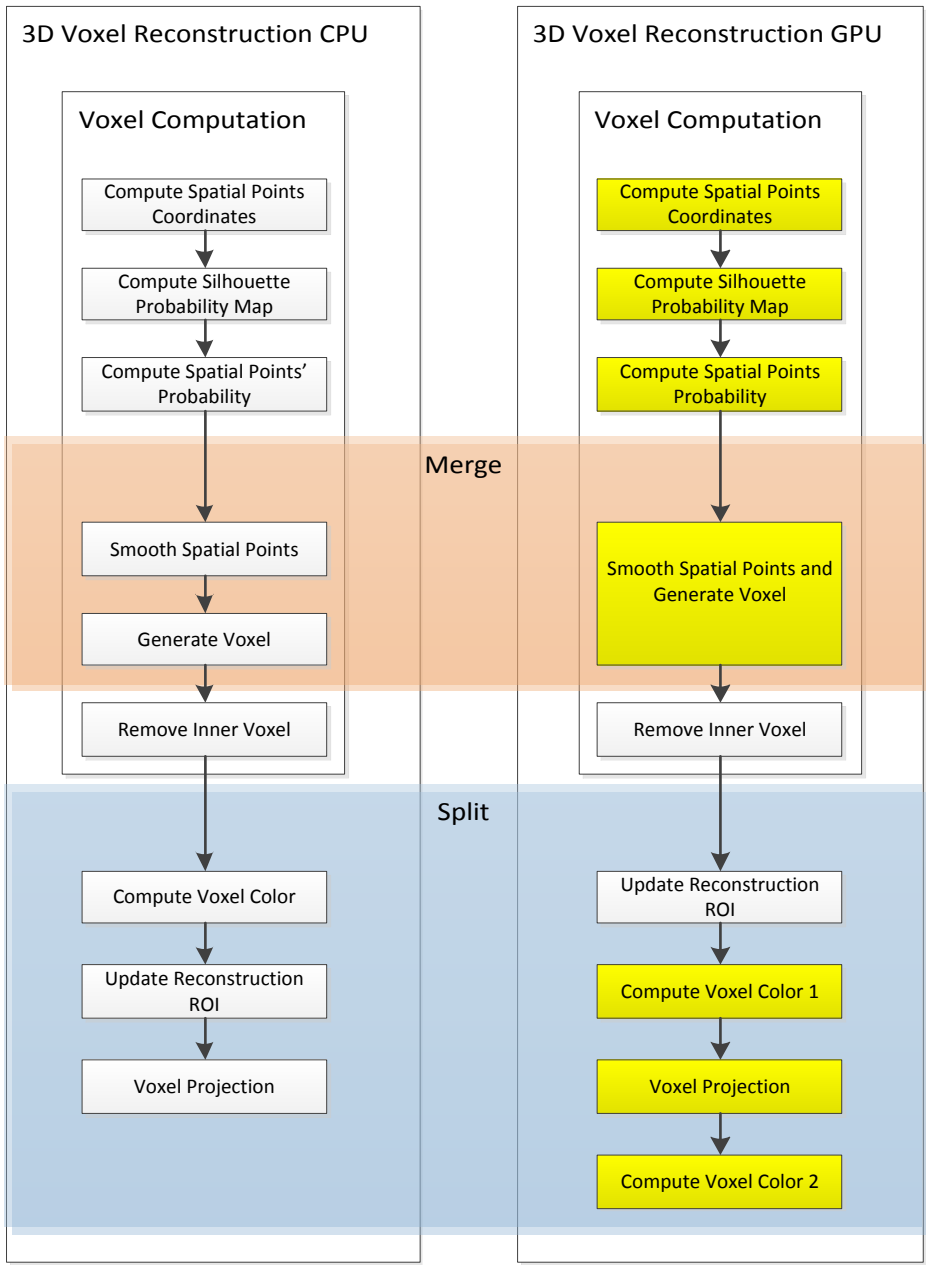

Figure 6 Comparison of CPU and GPU frameworks for 3-D voxel reconstruction algorithm

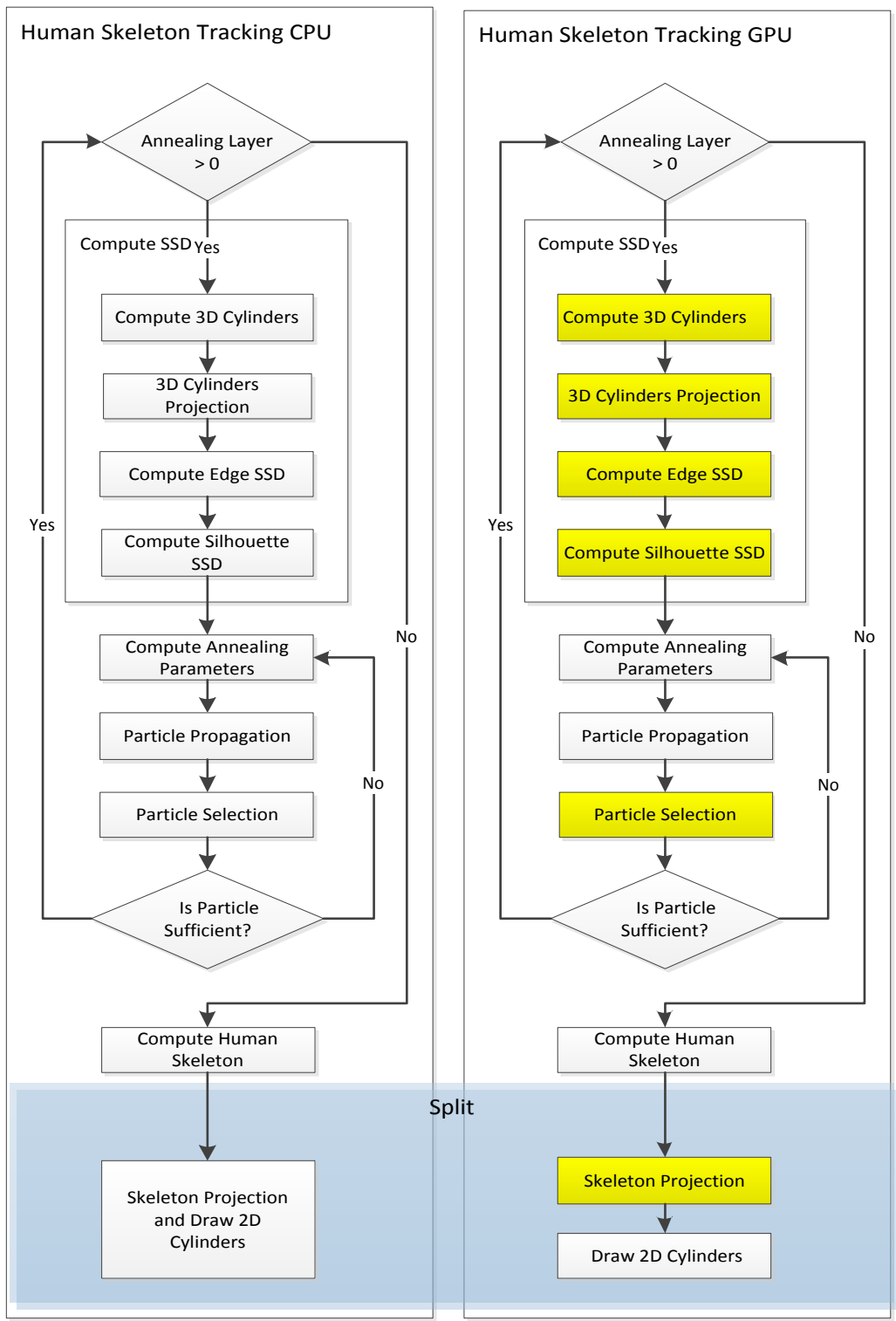

Figure 7 Comparison of CPU and GPU frameworks for human skeleton tracking algorithm

The comparisons of CPU and GPU algorithms are shown in Figure 6 and Figure 7, in which yellow background highlights GPU computing parts. For the parts which are more suitable for CPU processing, we modify interfaces and retain CPU implementation to achieve the overall optimum.

As shown in the figures, the algorithm has three major structural changes, that is, one merger and two splits.

- Merging Spatial Points Smoothing and Voxel Generating

In the spatial smoothing process, a float array with maximum length of 2250000 that is 8.6 MB of data is traversed and a new array with the same size is generated which is traversed later to generate filtered voxel cloud. Time complexity of each array traversing is same and memory copy is cheap when doing this on CPU. But for the sake of global memory access limitation on GPU, additional write and read operation of all the data in array will take much time when traversing the array twice other than once, so merging this procedure will significantly improve GPU computing efficiency.

- Splitting Voxel Color Computation

Voxel color calculation can be divided into two parts. Similar to the above analysis, the first half of which could share the same traverse process with voxel projection to reduce global memory access, and because of the different array lengths traversed in the first part and the second part, splitting will not increase memory access. So we split voxel computation into two parts handled by two kernels, respectively.

- Splitting Skeleton Projection and 2D Cylinders Drawing

In this step, the computed 3-D skeleton is projected to corresponding 2D planes of each camera and drawn to the images respectively. In CPU code, 2D cylinders on each image are drawn right after the end of computing skeleton projection in corresponding plane, while in GPU code, 2D cylinders on these images are not drawn until projection of all camera planes are completes. This is because projection computation can be parallelized on GPU and drawing on CPU does not spend lots of time. Moreover, drawing on CPU will refrain from two steps of large data copy.

### 3.3.2 Data Transfer Optimization

Data transfer always takes a long time in GPU program. We use the unblock way to read or write data to GPU device, and use multiple commandqueues to deal with different operations.

Read/write operation is arranged to the same commandqueue when host program needs to read/write data from device and the data will be used by the following kernel or CPU code. And if the data read/write from device is not needed immediately, the operation is put into another commandqueue, and the operations need these data wait for the events from read/write operation.

This will parallelize the CPU computation, GPU computation and data transfer as possible, to hide the time spent on CPU by serial code and the time spent on transferring data between host and device. The timeline after optimization is shown in Figure 8.

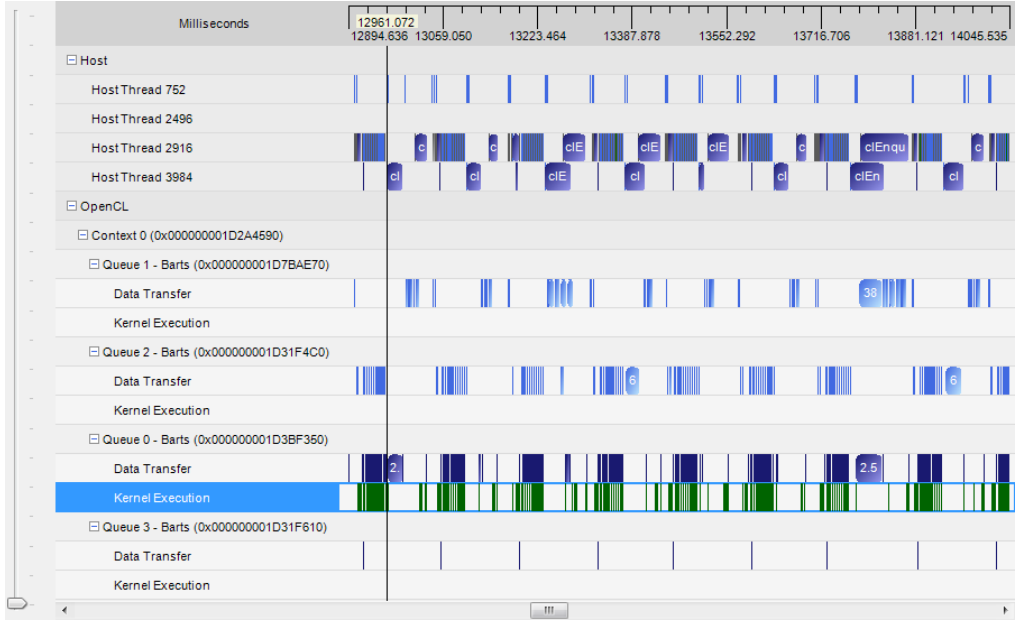

Figure 8 timeline after optimization

### 3.3.3 Main Program Optimization
**Voxel Computation**

As shown in Figure 6 this computation has 5 steps. Let *xlen* (maximum value is 150) denotes sample numbers of x coordinate, *ylen* (maximum value is 150) denotes sample numbers of y coordinate, *zlen* (maximum value is 100) denotes sample numbers of z coordinate, and *camNum* denotes camera number in use(it is set to be 8 in the program).

a) Compute spatial point coordinates. Compute spatial point coordinates using point location. Global work-item number is *xlen * ylen * zlen*, work-group size is set as *(4, 4, 4)*.
b) Compute silhouette probability map. It is computed separately for each camera, global work item number is image's pixel number.
c) Compute spatial points' probability. Compute points' probabilities using foreground probabilistic graph. The computation result for each camera is reduced using local memory. Global work-item number is *xlen * ylen * zlen * camNum*. Work-group size is *(256 / camNum, camNum)* and work-group number is *(ceil(xlen * ylen * zlen / (256 / camNum)), 1)*.
d) Smooth spatial points and generate voxel. Global work-item number is *xlen * ylen * zlen*, and work-group size is *(4, 4, 4)*.
e) Remove inner voxel. About 30000 numbers of surface voxel of human is sifted from a large amount of data with size of *xlen * ylen * zlen * 8*, which can be up to 18000000. This process is time-consuming on GPU so CPU implementation is reserved.

**Voxel Projection and Color Computation**

To make full use of GPU resources, we changed the code structure of color computation. Because of the optimization of algorithm implementation, we reduced time complexity without disturbing the correctness of the procedure. So this part of algorithm has a great speed up after the whole optimization on GPU. As shown in Figure 6, procedures in CPU program are split and disordered into 3 steps after optimization. The global work-item number is xlen * ylen * zlen *camNum in the first step and equals to image pixel number in the following two steps.

**Compute Sum-squared Difference (SSD)**

This part takes most of the time in skeleton tracking algorithm and has the most parallelism as well. As shown in Figure 7, it has 4 specific steps. Let *cylNum* denotes cylinder number (10 in the program), *camNum* denotes camera number (8 in the program), *particeNum* denotes particle number (can be adjusted according to different data, 200 for test data).

a) Compute Spatial 3-D Cylinders. Ten cylinders' spatial locations and orientations are computed according to the 31 parameters of each particle. The global work-item number is set as *particleNum*.
b) 3-D Cylinders Projection. Project spatial 3-D cylinders to the corresponding 2D plane of each camera. Global work-item number is *particleNum * camNum * cylNum*. One specific work-group is in charge of computation of one particle, work-group size is *(camNum, cylNum)*, and work-group number is *(particleNum, 1)*.
c) Compute Edge SSD. Edge SSD is computed using cylinder projections on 2D planes and edge maps. Global work-item number is set as *particleNum * camNum * cylNum * 2*. The computation of one specific particle lies in one work-group, and then data reducing is issued using local memory. The work-group size is *(camNum, cylNum, 2)*, and work-group number is *(particleNum, 1, 1)*.
d) Compute Silhouette SSD. Silhouette SSD value is computed using cylinder projections on 2D planes and silhouette maps. Global work-item number is *particleNum * camNum * cylNum*. One specific work-group is in charge of computation of one particle, local memory is used to reduce data. The work-group size is *(camNum, cylNum)*, and work-group number is *(particleNum, 1)*.

## 3.4  Parallelization Result and Performance Analysis

- IDE

    Microsoft Visual Studio 2008, 32 bit MSVC compiler

- 3rd Party Library
    OpenCV 2.3RC

    QT Library 4.7.3

    Intel TBB3.0 update 7

- Test Environment
    CPU: Intel(R) Core(TM) i7 CPU 860 @ 2.80GHz

    RAM: 4.00G DDR3

    GPU: AMD Radeon HD 6870

    Driver date 2011/7/7, version 8.872.0.0

    OS: Windows 7 x64 Enterprise SP1

    OpenCL 1.1 AMD-APP-SDK-v2.5 (684.213)

CAL 1.4.1457 (VM)

- Test Data and Algorithm Parameters
  Data: PEAR Dataset, Subject1_Walk1_Validate, 200 frames

  Particle Number: 200

  Annealed Layer Number: 10

We have implemented the CPU version, CPU + TBB parallel version and GPU version of algorithms, and following paragraphs compare these algorithms to evaluate the performance of GPU acceleration.

Because of the algorithm overlapping in optimization, we are not able to evaluate our algorithms individually. The following analysis takes the whole algorithm procedures into account, including voxel reconstruction, voxel color computation and human skeleton tracking.

TBB is a tool similar to OpenMP, the algorithm paralleled with TBB is essentially the same with the one not using TBB. The only difference is that TBB will automatically create numbers of threads when calculating for loops according to different CPU core number in runtime. Acceleration is achieved by using all available CPU cores. TBB is not used in the GPU accelerated algorithm, as all parallelism is performed on GPU.

In the code submitted in preliminary contest, the default algorithm running on CPU is the CPU+TBB version, which is paralleled using TBB. And due to the sake of automatically assigning threads according to the numbers of CPU cores in TBB, the runtime speed can change exponentially in this version of program when using different CPUs, especially those with different numbers of cores. This time change is not linear because not all parts of the algorithm are calculated using TBB. We use a 4-core Intel CPU to test the TBB-paralleled algorithm, and the result is used to illustrate relative acceleration of GPU program.

|  | Time 1(ms) | Time 2(ms) | Time 3(ms) | Average (ms) |
|---|---|---|---|---|
| CPU | 33948.37 | 33924.86 | 33759.48 | 33877.57 |
| CPU+TBB (4 cores) | 9605.077 | 9597.044 | 9593.483 | 9598.535 |
| GPU(pre) | 219.8729 | 217.0209 | 218.3901 | 218.428 |
| GPU(final) | 84.53274 | 85.58835 | 84.41453 | 84.845207 |

Table 2 average time per frame (results of 3 times of running)

Table 2 shows the average time consumed per frame of the algorithm running a total 200 frames. The table lists the result of running the algorithm for 3 times as well as the average value. Time consumed on GPU has two versions which represent the time of previously submitted program and the final program. More detailed values of the time cost by each frame can be found in Table 4 in the appendix.

The algorithm takes an average of more than 30 seconds per frame on CPU, while the GPU speed up to about 200 milliseconds per frame, which is close to real time. For scenes with non-violent movements (such as walking), the speed of 200 milliseconds per frame or 5 fps is rather a real-time performance.

The specific acceleration ratios can be computed from Table 2, which are listed in Table 3.

|  | Acceleration Ratio (pre) | Acceleration Ratio (final) |
| --- | --- | --- |
| CPU+TBB (4 cores) / CPU | 3.5295 | 3.5295 |
| GPU / CPU+TBB (4 cores) | 43.9437 | 113.1300 |
| GPU/ CPU | 155.0972 | 399.2868 |

Table 3 Acceleration ration of the algorithm

CPU program parallelized using TBB on 4-core CPU has an acceleration ratio of 3.5, which varies with CPU core numbers.

In the previously submitted program, the acceleration ratio of GPU program compared to CPU program parallelized using TBB on 4-core CPU is a speed up of 44 times, varies with CPU core numbers as well. And the GPU program speeds up about 155 times compared to the original CPU program without TBB parallelism.

The final version of algorithm has an acceleration ratio of GPU program compared to CPU program parallelized using TBB on 4-core CPU of 44 times, varies with CPU core numbers as well.

The final GPU program speeds up about 399 times compared to the original CPU program without TBB parallelism.

## 4    Prospect

Kinect, a latest product brought out by Microsoft (refer [8] for algorithm details) is one of the paragons for the commercialization of such technique. On the other hand, the startup companies like the "4D View Solution" as mentioned in the previous context are trying to provide the end to end solutions to customers, which to some extent prompt its application in animation, communication, sports and surveillance and such. Based on the existing solutions and methods, one can observe that this technology exhibits a bright prospect based on a solid theory, while the high performance computing platform is still under its infancy. We have tested our reconstructing and tracking algorithm on PEAR database, which originally consumed a lot of time and now has been accelerated to real-time via GPU. Our algorithm running on multi-camera systems or Kienct systems can play a role in animation, communication, surveillance and such areas.

## References

[1] Junchi Yan, Yin Li, Enliang Zheng, Yuncai Liu: An accelerated human motion tracking system based on voxel reconstruction under complex environments. Asian Conference on Computer Vision. (2009)

[2] Alexandru O. Balan, Leonid Sigal, Michael J. Black: A quantitative evaluation of video-based 3D person tracking. ICCCN '05 Proceedings of the 14th International Conference on Computer Communications and Networks.

[3] Jan-Michael Frahm, Pierre Georgel, David Gallup, Tim Johnson, Rahul Raguram, Changchang Wu, Yi-Hung Jen, Enrique Dunn, Brian Clipp Svetlana Lazebnik, Marc Pollefeys: Building Rome on a Cloudless Day. European Conference on Computer Vision. (2010)

[4] Agarwal, S., Snavely, N., Simon, I., Seitz, S.M., Szeliski, R.: Building Rome in a day. In: ICCV.

(2009)

[5] Sofiane Yous, Hamid Laga, Masatsugu Kidode, Kunihiro Chihara: GPU-Based Shape from Silhouettes. Proceedings of the 5th international conference on Computer graphics and interactive techniques in Australia and Southeast Asia. (2007)

[6] Neal Orman, Hansung Kim, Ryuuki Sakamoto, Tomoji Toriyama, Kiyoshi Kogure, and Robert Lindeman: GPU-Based Optimization of a Free-Viewpoint Video System. Electronic Letters on Computer Vision and Image Analysis 7(2):120-133, 2008.

[7] Franco, J., Boyer, E.: Fusion of multi-View silhouette cues using a space occupancy grid. In: Proc. of ICCV 2005, Beijing, China, vol. 1, pp. 1747–1754 (2005)

[8] Jamie Shotton, Andrew Fitzgibbon, Mat Cook, Toby Sharp, Mark Finocchio, Richard Moore, Alex Kipman and Andrew Blake: Real-Time Human Pose Recognition in Parts from Single Depth Images. Conference on Computer Vision and Pattern Recognition. (2011) CVPR'11 best paper award.

[9] K. Kutulakos and S. Seitz. A theory of shape by space carving. IJCV, 38(3):199–218, 2000.

[10] B. Goldluecke, M. Magnor. Space-time isosurface evolution for temporally coherent 3D reconstruction. CVPR2004.

[11] G. Vogiatzis, PHS. Torr, R. Cipolla. Multi-view stereo via volumetric graph-cuts. CVPR2005.

[12] G. Cheung, T. Kanade, JY. Bouguet, M. Holler. A real time system for robust 3D voxel reconstruction of human motions. CVPR2000.

[13] Laurentini, A.: The visual hull concept for silhouette- based image understanding. PAMI 16 (1994)

[14] J. Duetscher, A. Blake, I. Reid, "Articulated Body Motion Capture by Annealed Particle Filtering," cvpr, vol. 2, pp.2126, 2000 IEEE Computer Society Conference on Computer Vision and Pattern Recognition (CVPR'00) - Volume 2, 2000

## Appendix

Test results of running each algorithm 3 times.

| frame | CPU (ms) | | | CPU + TBB (4 cores) (ms) | | | GPU (ms) | | |
|---|---|---|---|---|---|---|---|---|---|
| | 1 | 2 | 3 | 1 | 2 | 3 | 1 | 2 | 3 |
| 1 | 35600.5 | 35652.8 | 35513.9 | 10252 | 10255.6 | 10359.3 | 145.425 | 152.239 | 155.904 |
| 2 | 33641.8 | 33700.1 | 33526 | 9470.22 | 9486.45 | 9465.98 | 81.5151 | 80.5634 | 125.687 |
| 3 | 32839.9 | 32920.3 | 32742.7 | 9219.91 | 9252.42 | 9213.83 | 103.1 | 80.5452 | 87.0664 |
| 4 | 32906.7 | 32995 | 32810 | 9126.02 | 9112.35 | 9274.83 | 82.0401 | 70.1632 | 84.3146 |
| 5 | 33318.4 | 33375.6 | 33204.6 | 9325.89 | 9269.03 | 9330.55 | 68.9786 | 72.0247 | 72.4673 |
| 6 | 33937.3 | 33993.8 | 33826 | 9443.41 | 9358.85 | 9341.25 | 95.623 | 79.8255 | 97.72 |
| 7 | 34964.3 | 34996.6 | 34828.6 | 9556.24 | 9535.68 | 9492.39 | 86.9925 | 71.6262 | 119.297 |
| 8 | 34794.8 | 34836.6 | 34660.4 | 9503.72 | 9474.04 | 9622.65 | 58.1777 | 75.3865 | 137.875 |
| 9 | 34388.3 | 34414.5 | 34260.6 | 9409.42 | 9324.38 | 9356.72 | 79.9194 | 89.0019 | 82.4673 |
| 10 | 36161.4 | 36156.4 | 36014.8 | 9802.5 | 9958.93 | 9822.64 | 90.3793 | 92.695 | 79.4183 |
| 11 | 36471.5 | 36427.7 | 36290.7 | 9972.38 | 9936.55 | 9838.54 | 124.304 | 78.8492 | 103.324 |
| 12 | 34901.2 | 34903.5 | 34751.5 | 9450.02 | 9569.22 | 9540.04 | 79.0944 | 93.1318 | 77.7996 |
| 13 | 34684.8 | 34648.2 | 34480.4 | 9455.61 | 9445.71 | 9437.74 | 88.0074 | 94.6182 | 104.53 |
| 14 | 34504 | 34501.8 | 34335.1 | 9462.61 | 9421.62 | 9450.51 | 98.1382 | 80.0257 | 77.1069 |
| 15 | 33844.9 | 33832.5 | 33656.4 | 9305.92 | 9358.47 | 9362.19 | 111.705 | 76.4321 | 69.0757 |
| 16 | 32626.2 | 32625.8 | 32473.8 | 9057.59 | 8940.46 | 9098.72 | 114.074 | 82.9919 | 72.6897 |
| 17 | 32540.9 | 32532.1 | 32377.7 | 9027.06 | 9039.79 | 9022.08 | 62.7196 | 78.6232 | 106.246 |
| 18 | 32120.2 | 32021.1 | 31893.9 | 8907.06 | 8838.18 | 9029.99 | 76.5956 | 78.7993 | 76.1069 |
| 19 | 32425.7 | 32418.7 | 32265 | 9072.47 | 9186.75 | 9085.62 | 94.2722 | 74.8075 | 78.3129 |
| 20 | 33669.3 | 33657.4 | 33514 | 9342.12 | 9366.85 | 9448.53 | 71.1893 | 78.3585 | 78.8299 |
| 21 | 34603.9 | 34613.4 | 34451 | 9698.33 | 9789.87 | 9783.41 | 101.314 | 118.421 | 81.2375 |
| 22 | 35053.4 | 35039.6 | 34901 | 9815.35 | 9844.04 | 9747.45 | 90.0716 | 75.039 | 85.0408 |
| 23 | 34836.2 | 34859.5 | 34708.4 | 9803.59 | 9697.96 | 9719.41 | 67.96 | 83.2522 | 74.0328 |
| 24 | 34553.8 | 34566.8 | 34405.4 | 9648.14 | 9711.33 | 9743.52 | 83.5859 | 104.65 | 104.85 |
| 25 | 35706 | 35681.5 | 35536.8 | 10049.4 | 10070 | 9968.84 | 106.472 | 71.4232 | 114.139 |
| 26 | 34915.8 | 34904.8 | 34767.2 | 9960.92 | 9962.03 | 9759.04 | 77.8377 | 111.941 | 75.2556 |
| 27 | 34396.4 | 34386 | 34231.8 | 9655.25 | 9788.04 | 9785.4 | 59.7662 | 79.7767 | 100.407 |
| 28 | 33520.3 | 33520.9 | 33349.2 | 9573.41 | 9495.59 | 9551.05 | 80.724 | 75.1254 | 77.7019 |
| 29 | 32804 | 32794.6 | 32639.3 | 9376.93 | 9520.01 | 9582.51 | 90.9878 | 78.1525 | 79.1405 |
| 30 | 32780.7 | 32776.9 | 32594.6 | 9413.95 | 9441.96 | 9548.58 | 78.7108 | 110.502 | 78.8973 |
| 31 | 32763.6 | 32770.5 | 32609.8 | 9335.42 | 9400.15 | 9461.49 | 121.127 | 72.0608 | 75.6417 |
| 32 | 31668.6 | 31662 | 31515.8 | 9126.74 | 9224.86 | 9081.39 | 74.4544 | 80.1044 | 89.6454 |
| 33 | 31059.6 | 31048.4 | 30890.6 | 8929.74 | 8992.01 | 9005.34 | 78.487 | 75.077 | 81.1716 |
| 34 | 31227.6 | 31237 | 31064.8 | 9125.55 | 9133.15 | 9021.72 | 71.6378 | 79.8033 | 72.3871 |
| 35 | 30276.6 | 30288.6 | 30127.3 | 8765.8 | 8845.49 | 8767.27 | 59.6867 | 75.9556 | 68.8992 |
| 36 | 30735.7 | 30736.8 | 30575.8 | 8816.05 | 8888.65 | 8990.87 | 76.2438 | 55.8981 | 62.8557 |
| 37 | 31162.1 | 31135.4 | 30976 | 8861.47 | 9134.62 | 9491.66 | 90.3319 | 74.7857 | 73.2792 |
| 38 | 31974.9 | 31987.3 | 31833.8 | 9268.02 | 9486.73 | 9512.07 | 69.9381 | 77.7708 | 74.2322 |
| 39 | 31785.2 | 31773.7 | 31613.6 | 9347.38 | 9339.87 | 9366.08 | 78.7553 | 73.9125 | 115.611 |
| 40 | 33316.4 | 33318.4 | 33153 | 9484.22 | 9594.7 | 9647.29 | 72.7781 | 83.8038 | 69.5103 |
| 41 | 33031.3 | 33018.4 | 32860.7 | 9672.67 | 9693.39 | 9585.59 | 65.7062 | 77.7194 | 109.873 |
| 42 | 32427.5 | 32432 | 32280.9 | 9291.26 | 9108.74 | 9395.02 | 72.2306 | 108.33 | 70.5912 |
| 43 | 33426.2 | 33405.4 | 33254.2 | 9513.36 | 9478.82 | 9565.69 | 85.0861 | 69.3364 | 79.7916 |
| 44 | 34943.7 | 34937 | 34753.2 | 9843.74 | 9799.12 | 9978.72 | 91.9732 | 85.8407 | 117.604 |
| 45 | 36321.3 | 36315.5 | 36145 | 10286.6 | 10393 | 10197.7 | 96.4666 | 85.0218 | 68.9783 |
| 46 | 38029.6 | 38015.1 | 37876.8 | 10882 | 10650.4 | 10737.1 | 84.0445 | 73.2399 | 93.931 |
| 47 | 38452.1 | 38461 | 38284.3 | 10926.6 | 10795.9 | 10975.3 | 75.7749 | 72.3787 | 80.1562 |
| 48 | 38847.4 | 38826.7 | 38659.9 | 10757.8 | 10671.3 | 10713.3 | 73.7413 | 76.8528 | 84.3612 |
| 49 | 37603 | 37577.1 | 37422.5 | 10521.1 | 10553.6 | 10222.9 | 80.4699 | 88.5739 | 70.9303 |
| 50 | 37000.4 | 36992.7 | 36811.2 | 10353.3 | 10405.6 | 10385.3 | 79.9208 | 72.9923 | 71.4615 |
| 51 | 35987.8 | 35943.1 | 35770.6 | 10052.3 | 10048.2 | 10031.5 | 126.185 | 73.343 | 58.6751 |
| 52 | 34258.9 | 34224.1 | 34067.4 | 9601.21 | 9604.71 | 9561.01 | 85.3307 | 84.1939 | 82.6043 |
| 53 | 33064.5 | 33200.9 | 33025.5 | 9192.81 | 9263.89 | 9296.56 | 74.27 | 69.8468 | 84.5632 |
| 54 | 32897.2 | 33057.8 | 32922.4 | 9328.34 | 9240.47 | 9194.8 | 75.7702 | 104.494 | 109.212 |
| 55 | 33925 | 33887.9 | 33747.7 | 9447.58 | 9466.66 | 9402.81 | 74.4213 | 120.224 | 115.111 |
| 56 | 33860.7 | 33868.4 | 33713.3 | 9544.03 | 9552.05 | 9631.64 | 75.8697 | 94.4497 | 67.0709 |
| 57 | 34682.3 | 34659.9 | 34511 | 9874.57 | 9599.09 | 9825.64 | 120.497 | 81.7706 | 73.4324 |
| 58 | 35191.2 | 35158.3 | 35006.8 | 10185.1 | 10192.8 | 9988.62 | 119.73 | 81.0111 | 99.4036 |
| 59 | 34667.1 | 34612.1 | 34474.2 | 9993.62 | 9750.95 | 9736.79 | 109.874 | 66.6899 | 71.0681 |
| 60 | 33222.2 | 33191.5 | 33054.5 | 9208.42 | 9709.71 | 9485.01 | 83.9497 | 123.989 | 73.1895 |
| 61 | 32288.1 | 32267 | 32104.5 | 9313.55 | 9427.36 | 9300.34 | 95.3532 | 80.9776 | 56.2674 |
| 62 | 32492 | 32484.3 | 32308.7 | 9615.77 | 9542.18 | 9367.01 | 70.4252 | 101.233 | 93.1894 |
| 63 | 32550.5 | 32528.6 | 32371.4 | 9569.48 | 9296.49 | 9886.95 | 106.304 | 83.9963 | 85.1148 |
| 64 | 32507.7 | 32506.6 | 32357 | 9506.64 | 9553.2 | 9586.39 | 91.4752 | 75.9286 | 75.6708 |
| 65 | 32372.1 | 32367.6 | 32215.1 | 9535.09 | 9600.36 | 9392.11 | 89.3042 | 107.808 | 103.626 |
| 66 | 31496.4 | 31481.9 | 31333.2 | 8967.79 | 9122.01 | 9225.35 | 79.5738 | 79.9714 | 65.719 |
| 67 | 32481.2 | 32464 | 32305.3 | 9508.83 | 9569.94 | 9719.06 | 76.0145 | 97.7521 | 78.5882 |
| 68 | 33476.2 | 33459.6 | 33296.1 | 9577.65 | 9859.55 | 9828.11 | 72.0151 | 75.5104 | 103.743 |
| 69 | 33203.5 | 33181.2 | 33032.6 | 9796.9 | 9942.04 | 9582.99 | 96.0838 | 91.096 | 81.0757 |
| 70 | 31828.2 | 31793.4 | 31657.8 | 9240.18 | 9558.47 | 9267.62 | 91.6349 | 72.7586 | 72.5997 |
| 71 | 31188.2 | 31156.2 | 31002.1 | 9071.39 | 9280.09 | 9146.51 | 80.2595 | 105.284 | 72.3164 |
| 72 | 31683.3 | 31498.1 | 31360.1 | 9373.42 | 9127.2 | 9265.59 | 81.8104 | 116.5 | 70.2492 |
| 73 | 32543.2 | 32512.5 | 32387.6 | 9480.85 | 9372.52 | 9251.81 | 59.0028 | 73.5005 | 73.9734 |
| 74 | 32096.2 | 32066.7 | 31959.5 | 9219.26 | 9218.76 | 9449.79 | 78.9095 | 77.4059 | 70.4057 |
| 75 | 31511.5 | 31465.5 | 31314.4 | 9188.55 | 9120.31 | 9111.47 | 80.8563 | 99.5173 | 73.7051 |
| 76 | 31865.3 | 31853.8 | 31709.4 | 9233.38 | 9089.69 | 9268.59 | 83.5006 | 77.5593 | 69.737 |
| 77 | 32979.6 | 32977.2 | 32870.7 | 9544.74 | 9574.74 | 9562.96 | 74.8318 | 86.6898 | 90.8937 |
| 78 | 32667.4 | 32611.1 | 32494.1 | 9309.08 | 9314.98 | 9212.08 | 65.6737 | 102.47 | 79.061 |
| 79 | 32960.5 | 32892 | 32784.8 | 9624.06 | 9317.88 | 9397.76 | 85.8338 | 83.3021 | 125.238 |
| 80 | 34155 | 34125.9 | 33956.1 | 9656.97 | 9860.35 | 9630.47 | 86.5794 | 90.0931 | 67.9532 |
| 81 | 33706.4 | 33658.1 | 33531.2 | 9457.28 | 9705.61 | 9539.47 | 73.8343 | 92.6349 | 96.7703 |
| 82 | 34551.3 | 34510 | 34352.7 | 9707.29 | 9801.87 | 9812.67 | 75.3123 | 82.1235 | 79.3531 |
| 83 | 32927.6 | 32876.7 | 32741.9 | 9274.47 | 9430.79 | 9375.85 | 68.5239 | 59.4327 | 96.0615 |
| 84 | 31502.1 | 31447.6 | 31300 | 9034.32 | 9060.03 | 9004.95 | 85.6745 | 89.0974 | 61.2588 |
| 85 | 33020.6 | 33001 | 32834.1 | 9526.25 | 9363.36 | 9366.82 | 77.0768 | 82.5249 | 76.0012 |
| 86 | 33744.3 | 33623.8 | 33504.4 | 9579.51 | 9608.37 | 9672.01 | 120.279 | 75.2046 | 85.1272 |
| 87 | 33410.1 | 33404.4 | 33241.7 | 9415.11 | 9353.66 | 9495.11 | 72.4041 | 111.09 | 73.8079 |
| 88 | 34737.1 | 34689.2 | 34540.7 | 9775.95 | 9733.12 | 9750.64 | 91.0257 | 83.0072 | 72.7736 |
| 89 | 37060.8 | 37032.5 | 36888.3 | 10346 | 10583.8 | 10555.7 | 72.3589 | 80.6298 | 63.9352 |
| 90 | 37341 | 37301.4 | 37141 | 10330.6 | 10348.6 | 10250.7 | 71.4989 | 100.419 | 126.345 |
| 91 | 37261.7 | 37229.8 | 37109 | 10144.2 | 10332.2 | 10283.2 | 87.283 | 83.3692 | 78.6126 |
| 92 | 38923.4 | 38812.9 | 38656.4 | 10596.5 | 10651.8 | 10642.5 | 77.6095 | 83.2547 | 73.588 |
| 93 | 38461.8 | 38426.5 | 38286.8 | 10387.4 | 10514.7 | 10357.6 | 68.2553 | 74.6121 | 84.4276 |
| 94 | 37633.3 | 37600.7 | 37450.5 | 10221.6 | 10101.6 | 10127.8 | 70.5491 | 100.51 | 71.4571 |
| 95 | 38223.1 | 38183.3 | 38032.4 | 10265.6 | 10355.6 | 10359 | 86.0023 | 88.0482 | 86.6861 |
| 96 | 38832.4 | 38768.3 | 38600.1 | 10585.9 | 10517.5 | 10584.2 | 84.1626 | 104.914 | 82.5201 |
| 97 | 35965.8 | 35926.4 | 35784.8 | 9949.36 | 9963.97 | 9970.05 | 89.7399 | 114.61 | 75.6537 |
| 98 | 34907.1 | 34878.9 | 34707.1 | 9714.16 | 9505.28 | 9699.62 | 87.3428 | 85.7419 | 78.7934 |
| 99 | 33704.5 | 33637 | 33480.2 | 9461.26 | 9365.03 | 9382.82 | 82.0485 | 90.3793 | 92.8369 |
| 100 | 32883.6 | 32852.7 | 32686.9 | 9256.75 | 9154.32 | 9135.54 | 77.4516 | 60.055 | 90.2535 |
| 101 | 32465.6 | 32422.8 | 32292 | 9264.42 | 9213.85 | 9194.33 | 70.7562 | 81.1191 | 66.0676 |
| 102 | 32613 | 32579.9 | 32423.1 | 9196.7 | 9229.97 | 9214.47 | 102.307 | 128.318 | 70.9208 |
| 103 | 32291.4 | 32250.7 | 32115.1 | 9205.03 | 9051.55 | 9133.48 | 71.1966 | 77.1565 | 108.533 |
| 104 | 32609.8 | 32543.7 | 32395.4 | 9171.69 | 9201.56 | 9144.44 | 112.728 | 70.7742 | 62.1663 |
| 105 | 33079.6 | 32969.7 | 32847.3 | 9229.9 | 9170.76 | 9269.6 | 82.3781 | 86.0776 | 74.9796 |
| 106 | 33293.2 | 33278.3 | 33119.1 | 9323.55 | 9320.93 | 9358.31 | 72.9601 | 102.069 | 71.0962 |
| 107 | 32607.4 | 32539.5 | 32384.1 | 9047.67 | 9253.64 | 8946.29 | 72.4551 | 96.0951 | 69.9521 |
| 108 | 31920 | 31884 | 31718.7 | 9061.82 | 9002.48 | 9125.16 | 79.7896 | 81.2776 | 70.3561 |
| 109 | 30527.7 | 30511 | 30368.3 | 8518.69 | 8659.65 | 8542.84 | 140.959 | 80.939 | 76.6727 |
| 110 | 31119.8 | 31110.8 | 30941 | 8830.54 | 8800.78 | 8760.15 | 104.741 | 117.999 | 73.3817 |
| 111 | 31180.4 | 31158.6 | 31005.9 | 8842.46 | 8733.13 | 8709.92 | 81.2049 | 70.6025 | 61.3215 |
| 112 | 30995.6 | 30955.7 | 30790.2 | 8604.25 | 8731.63 | 8756.32 | 79.0397 | 82.818 | 95.9952 |
| 113 | 31067.6 | 31031.4 | 30863.6 | 8848.25 | 8889.48 | 8787.71 | 130.641 | 104.967 | 80.2853 |
| 114 | 31028.9 | 31025.7 | 30839.8 | 8898.08 | 8813.58 | 8728.41 | 94.7039 | 69.566 | 69.8927 |
| 115 | 31152.1 | 31095.4 | 30913.4 | 8899.61 | 8909.59 | 8828.28 | 77.3925 | 105.244 | 116.389 |
| 116 | 31280.5 | 31251.5 | 31073.3 | 8967.47 | 8993.1 | 8987.81 | 77.8781 | 72.7014 | 75.6526 |
| 117 | 31773.5 | 31737.2 | 31560.5 | 9172.51 | 9152.94 | 9151.95 | 89.5528 | 77.7744 | 79.9852 |
| 118 | 31756 | 31735.1 | 31552 | 9179.07 | 9218.74 | 9181.63 | 75.6185 | 72.1913 | 77.279 |
| 119 | 31564.8 | 31516.9 | 31342.7 | 9099.27 | 9172.31 | 9121.55 | 67.7711 | 72.8709 | 123.153 |
| 120 | 32980.4 | 32945.7 | 32781.9 | 9782.39 | 9628.33 | 9552.71 | 78.0772 | 63.4178 | 70.7473 |
| 121 | 32295.7 | 32263.6 | 32116.8 | 9495.12 | 9451.08 | 9618.83 | 71.8674 | 78.784 | 76.5287 |
| 122 | 32614.5 | 32586.9 | 32436.8 | 9548.09 | 9654.34 | 9479.96 | 111.471 | 106.218 | 73.3011 |
| 123 | 32611.5 | 32572.1 | 32419.8 | 9755.21 | 9660 | 9497.57 | 75.2361 | 81.4275 | 83.9423 |
| 124 | 32493.6 | 32450.6 | 32270.9 | 9491.3 | 9619.31 | 9341.97 | 58.3546 | 66.6017 | 70.9562 |
| 125 | 32381.3 | 32323.4 | 32155.9 | 9685.26 | 9671.44 | 9678.93 | 80.1297 | 77.4503 | 70.3809 |
| 126 | 31769.3 | 31728.2 | 31558.1 | 9567.76 | 9465.09 | 9649.87 | 83.5385 | 110.177 | 80.9889 |
| 127 | 32645.2 | 32604.2 | 32427.8 | 9594.68 | 9605.59 | 9594.27 | 87.9684 | 137.323 | 75.9928 |
| 128 | 32637.7 | 32610.6 | 32424.1 | 9271.57 | 9544.22 | 9594.95 | 72.9083 | 59.5581 | 80.2649 |
| 129 | 32830.1 | 32790.4 | 32609.9 | 9362.45 | 9870.55 | 9500.13 | 68.6453 | 86.1301 | 77.9677 |
| 130 | 32288.6 | 32297.2 | 32098.9 | 9144.07 | 9663.86 | 9377.78 | 78.1866 | 74.5082 | 72.4524 |
| 131 | 32927.9 | 32936.9 | 32712.2 | 9669.63 | 9492.67 | 9519.01 | 73.0888 | 75.2917 | 66.9214 |
| 132 | 33934 | 33839.2 | 33657.6 | 9436.02 | 9540.87 | 9776.29 | 81.0466 | 73.0969 | 81.0378 |
| 133 | 35150.2 | 35122.6 | 34960.5 | 10319.7 | 10056.3 | 9794.28 | 74.146 | 59.9033 | 83.4841 |
| 134 | 34581.3 | 34548.9 | 34379.8 | 9990.77 | 9720.92 | 9590.98 | 66.9756 | 82.9215 | 120.903 |
| 135 | 34787.3 | 34745.2 | 34594.3 | 10056.7 | 9672.89 | 9841.61 | 70.7683 | 76.6632 | 78.8222 |
| 136 | 34629.5 | 34642.9 | 34451.8 | 9589.41 | 9687.03 | 9480.16 | 74.7727 | 119.406 | 120.548 |
| 137 | 35615.7 | 35616.8 | 35413.5 | 10107.2 | 9696.28 | 9815.82 | 106.854 | 76.198 | 77.8816 |
| 138 | 35248 | 35235.9 | 35069.9 | 9798.69 | 10010.2 | 9950.97 | 77.5734 | 65.5659 | 75.5586 |
| 139 | 36444.2 | 36407.5 | 36264.4 | 9932.79 | 9954.49 | 10156.4 | 98.5388 | 73.1053 | 116.89 |
| 140 | 38475.7 | 38460.3 | 38315.8 | 10486.6 | 10345.8 | 10449.2 | 87.8878 | 71.6711 | 124.432 |
| 141 | 38826.3 | 38810.5 | 38641.2 | 10808.7 | 10684.2 | 11047.5 | 83.6898 | 76.0697 | 58.4757 |
| 142 | 38667.7 | 38664.8 | 38486.2 | 10699 | 10869.7 | 10830.7 | 75.3786 | 78.2389 | 87.8641 |
| 143 | 37816.8 | 37806.7 | 37602.2 | 10815.6 | 10560.6 | 10323.1 | 116.186 | 63.7769 | 79.201 |
| 144 | 36781.1 | 36781.3 | 36577.8 | 10342.3 | 10469.5 | 10506.8 | 80.0685 | 87.9742 | 76.6946 |
| 145 | 36391.6 | 36340.1 | 36114.9 | 10190 | 10012.8 | 10244.5 | 77.1224 | 101.829 | 103.296 |
| 146 | 35727.6 | 35696.5 | 35500.5 | 9840.16 | 10072 | 10055.2 | 65.6336 | 108.198 | 80.7672 |
| 147 | 35243.6 | 35241.7 | 35036.7 | 9944.68 | 10002.6 | 10175.2 | 81.5472 | 100.31 | 93.2302 |
| 148 | 34713.5 | 34704.9 | 34511.1 | 9712.35 | 9777.54 | 9840.58 | 94.2168 | 103.486 | 89.1178 |
| 149 | 35782.2 | 35739 | 35556.6 | 10267.6 | 9776.74 | 9879.22 | 116.739 | 79.3038 | 86.6274 |
| 150 | 36870.3 | 36829.5 | 36660.3 | 10381.4 | 10456.3 | 10260.4 | 82.0018 | 83.9606 | 78.9108 |
| 151 | 35884.3 | 35872.6 | 35668.5 | 9893.97 | 9919.35 | 9975.89 | 97.7659 | 89.0701 | 80.0159 |
| 152 | 35362.5 | 35320.3 | 35112.6 | 10082.2 | 9818.4 | 9828.3 | 78.1016 | 79.543 | 74.0784 |
| 153 | 35543.2 | 35573.5 | 35335.2 | 10181.6 | 10066.5 | 10062.8 | 81.2862 | 82.5712 | 72.3033 |
| 154 | 33883.7 | 33862.4 | 33671.5 | 9729.62 | 9581.49 | 9651.09 | 107.016 | 70.9569 | 106.169 |
| 155 | 34031.4 | 34006.4 | 33828.7 | 9655.93 | 9644.88 | 9618.74 | 104.75 | 83.1457 | 83.4334 |
| 156 | 34603.5 | 34585.7 | 34402 | 9789.67 | 9553.68 | 9785.04 | 135.744 | 62.4127 | 81.725 |
| 157 | 34177 | 34176.6 | 33996 | 9508.7 | 9409.59 | 9292.89 | 71.4183 | 74.8174 | 90.5608 |
| 158 | 33388.2 | 33377.8 | 33207.3 | 9275.76 | 9255.48 | 9025.49 | 73.8011 | 77.6775 | 72.8188 |
| 159 | 33793.3 | 33729.9 | 33564.8 | 9177.4 | 9411.85 | 9524.19 | 74.3086 | 87.6825 | 99.7467 |
| 160 | 33525 | 33527.5 | 33340.2 | 9346.66 | 9418.04 | 9181.44 | 90.7198 | 90.6618 | 95.3375 |
| 161 | 34252.3 | 34240.7 | 34065.6 | 9521.23 | 9332.23 | 9402.89 | 90.1114 | 101.626 | 77.9615 |
| 162 | 35154.7 | 35110.9 | 34925 | 9817.59 | 9692.39 | 9767.24 | 72.9386 | 87.8878 | 95.9361 |
| 163 | 34214.1 | 34212.5 | 34014.6 | 9523.71 | 9318.69 | 9387.01 | 71.543 | 80.4563 | 86.2268 |
| 164 | 35573.7 | 35568.8 | 35354.3 | 9751.06 | 9750.57 | 9704.39 | 79.4225 | 109.075 | 108.197 |
| 165 | 34800.4 | 34757.5 | 34530.4 | 9620.65 | 9609.85 | 9631.02 | 78.6664 | 79.4591 | 99.2155 |
| 166 | 34183 | 34180.9 | 33990.4 | 9534.36 | 9531.32 | 9570.82 | 76.678 | 67.7989 | 80.5824 |
| 167 | 34003.7 | 33977.3 | 33814.1 | 9598.1 | 9334.95 | 9508.76 | 64.1925 | 81.0425 | 77.2801 |
| 168 | 34070.4 | 34055.4 | 33877.4 | 9591.54 | 9536.41 | 9526.37 | 79.4436 | 87.1586 | 77.8809 |
| 169 | 33224.6 | 33208.6 | 33012.6 | 9220.73 | 9297.34 | 9322.78 | 87.5634 | 101.634 | 86.9373 |
| 170 | 33483 | 33358.4 | 33159.4 | 9431.6 | 9173.08 | 9189.25 | 88.6855 | 74.7361 | 75.4255 |
| 171 | 34615.1 | 34604.5 | 34441.3 | 9462.98 | 9628.39 | 9442.95 | 69.5262 | 58.8373 | 76.7077 |
| 172 | 34720.6 | 34686.2 | 34526.7 | 9594.76 | 9373.74 | 9397.32 | 73.2481 | 75.6333 | 111.174 |
| 173 | 34726.5 | 34683.2 | 34488.3 | 9687.58 | 9535.6 | 9602.04 | 73.7446 | 84.7765 | 70.912 |
| 174 | 35096 | 34925.1 | 34722.9 | 9512.11 | 9680.85 | 9558.86 | 126.667 | 73.0773 | 73.1513 |
| 175 | 35652.7 | 35624.7 | 35450.4 | 9912.82 | 9714.58 | 9776.43 | 74.8208 | 88.4762 | 61.9213 |
| 176 | 35404.3 | 35379.9 | 35191.3 | 9696.16 | 9693.73 | 9748.5 | 75.115 | 79.228 | 70.1406 |

| 177 | 35859 | 35827.6 | 35634.4 | 9978.46 | 9839.39 | 9910.85 | 120.921 | 81.963 | 71.0025 |
| --- | --- | --- | --- | --- | --- | --- | --- | --- | --- |
| 178 | 34434.8 | 34416.4 | 34213.4 | 9584.53 | 9633.77 | 9517.88 | 88.7891 | 111.574 | 82.7604 |
| 179 | 34040.8 | 33897.8 | 33707.6 | 9394.67 | 9291.61 | 9198.92 | 94.6704 | 78.392 | 75.8258 |
| 180 | 33710.7 | 33710.2 | 33514.2 | 9574.77 | 9442.56 | 9346.1 | 59.9033 | 76.7708 | 124.525 |
| 181 | 33249.5 | 33113.5 | 32925.1 | 9327.29 | 9335.31 | 9230.34 | 80.1228 | 119.917 | 89.8966 |
| 182 | 33442.2 | 33404.4 | 33220.5 | 9398.64 | 9317.44 | 9378.29 | 90.2083 | 75.4011 | 75.6344 |
| 183 | 34352.9 | 34271.5 | 34061.7 | 9746.24 | 9683.05 | 9832.89 | 81.895 | 126.98 | 109.449 |
| 184 | 34484.9 | 34452.1 | 34271 | 9844.01 | 9772.22 | 9716.8 | 94.3287 | 90.719 | 74.0656 |
| 185 | 34416.2 | 34320.7 | 34126.2 | 9641.81 | 9795.28 | 9718.13 | 122.258 | 105.543 | 79.1438 |
| 186 | 35908.5 | 35901.7 | 35701.1 | 10247.8 | 10124.1 | 10209.1 | 78.5996 | 72.4899 | 71.2387 |
| 187 | 35175 | 35131.1 | 34947.1 | 10088.8 | 10034.5 | 9962.42 | 110.584 | 83.1311 | 93.4676 |
| 188 | 35454.3 | 35376 | 35182.9 | 10008 | 9991.53 | 10107.4 | 79.292 | 101.74 | 71.3823 |
| 189 | 35021.7 | 34998.1 | 34801.8 | 10027.6 | 9870.78 | 9950.72 | 76.0287 | 76.295 | 101.45 |
| 190 | 33625.1 | 33574.9 | 33570.1 | 9522.63 | 9484.14 | 9531.37 | 61.9066 | 110.437 | 79.1051 |
| 191 | 33374.6 | 33324.3 | 33155.5 | 9501.46 | 9490.54 | 9551.4 | 69.0803 | 71.7706 | 81.9514 |
| 192 | 34185.4 | 34161.1 | 33978.4 | 9867.9 | 9601.94 | 9919.46 | 73.1044 | 118.437 | 75.7955 |
| 193 | 34063.9 | 34042.9 | 33847.3 | 9784.25 | 9837.59 | 9781.06 | 74.2521 | 73.7076 | 75.6847 |
| 194 | 32426.6 | 32406.5 | 32234.9 | 9359.52 | 9350.86 | 9416.13 | 65.6584 | 63.5487 | 72.7721 |
| 195 | 33487.6 | 33495 | 33331.3 | 9751.73 | 9793.88 | 9788.08 | 75.9995 | 92.547 | 119.086 |
| 196 | 34401.5 | 34372.5 | 34240.9 | 10012.2 | 9986.84 | 9762.91 | 120.223 | 74.6165 | 119.039 |
| 197 | 33440.9 | 33431.3 | 33246.5 | 9878.34 | 9724.13 | 9506.84 | 69.5229 | 86.3091 | 82.1687 |
| 198 | 32560.6 | 32561.7 | 32393.4 | 9553.09 | 9578.07 | 9457.74 | 63.9314 | 80.4358 | 81.1927 |
| 199 | 31543.9 | 31521.3 | 31320.8 | 9226.17 | 9274.09 | 9232.85 | 74.7483 | 71.0557 | 71.052 |
| 200 | 30689.9 | 30657.3 | 30520 | 9071.56 | 8989.1 | 8960.14 | 90.2441 | 125.848 | 103.54 |

Table 4 time of each frame (results of 3 times of running)